# New Evidence for a Low-Temperature Magnetic Ground State in Double-Perovskite Iridates with $Ir^{5+}(5d^4)$ Ions

J. Terizc[1], H. Zheng[1], Feng Ye[2,3], H. D. Zhao[1], P. Schlottmann[4], L. De Long[3], and G. Cao[1*]

[1]Department of Physics, University of Colorado-Boulder, Boulder, CO 80309

[2]Quantum Condensed Matter Division, Oak Ridge National Laboratory

Oak Ridge, TN 37831

[3]Department of Physics and Astronomy and Center for Advanced Materials

University of Kentucky, Lexington, KY 40506

[4]Department of Physics, Florida State University, Tallahassee, FL 32306

We report an unusual magnetic ground state in single-crystal, double-perovskite $Ba_2YIrO_6$ and Sr doped $Ba_2YIrO_6$ with $Ir^{5+}$ ($5d^4$) ions. Long-range magnetic order below 1.7 K is confirmed by DC magnetization, AC magnetic susceptibility and heat capacity measurements. The observed magnetic order is extraordinarily delicate and cannot be explained in terms of either a low-spin S=1 state, or a singlet $J_{eff} = 0$ state imposed by the spin-orbit interactions (SOI). Alternatively, the magnetic ground state appears consistent with a SOI that competes with comparable Hund's rule coupling and inherently large electron hopping, which cannot stabilize the singlet $J_{eff} = 0$ ground state. However, this picture is controversial, and conflicting magnetic behavior for these materials is reported in both experimental and theoretical studies, which highlights the intricate interplay of interactions that determine the ground state of materials with strong SOI.

**I. Introduction**

The physics of iridates is driven by a unique combination of strong spin-orbit interaction (SOI), large crystalline fields, and comparable onsite Coulomb interactions (U), which can create a delicate balance between interactions that drive exotic electronic states seldom observed in other materials. The seminal "$J_{eff} = 1/2$ Mott state" observed in layered iridates with tetravalent $Ir^{4+}(5d^5)$ ions is a profound manifestation of such circumstances [1-3]. A great deal of recent work has appeared in response to possible experimental observations and theoretical predictions of novel phenomena, including a large array of novel effects in 5d-electron systems having inherently strong SOI [4-23]. Most of these discussions have focused on tetravalent iridates such as $Sr_2IrO_4$, since the $Ir^{4+}(5d^5)$ ion provides four d-electrons occupying the lower $J_{eff} = 3/2$ bands, and one electron to partially fill the upper $J_{eff} = 1/2$ band that lies closest to the Fermi energy, and therefore dominates the new physics. One common occurrence in these spin-orbit coupled insulators is an antiferromagnetic (AFM) phase transition that often occurs at relatively high temperatures (e.g., at 240 K for $Sr_2IrO_4$ [24], 285 K for $Sr_3Ir_2O_7$ [25] and 185 K for orthorhombic perovskite $Sr_{0.94}Ir_{0.78}O_{2.68}$ [26]).

On the other hand, for iridates with pentavalent $Ir^{5+}(5d^4)$ ions, the strong SOI is expected to impose a nonmagnetic singlet ground state, $J_{eff} = 0$ [15, 27-36]. Very recent experimental and theoretical studies revealed strong evidence against the presence of a $J_{eff} = 0$ state [15, 28-32]. In particular, our previous experimental study concluded that the distorted double-perovskite $Sr_2YIrO_6$ exhibits an exotic magnetic state below 1.3 K which differs from the anticipated $J_{eff} = 0$ state [28]. The emergence of the exotic magnetic ground state was initially attributed to the effects of non-cubic crystal fields on the $J_{eff} = 1/2$ and $J_{eff}$ = 3/2 states [28], noting that such effects were not included in the original model [1, 2]. However, there are other possible mechanisms for stabilizing a magnetic moment in pentavalent Ir systems:

(1) The bandwidth of the $t_{2g}$ electrons increases with increased hopping, and the $J_{eff}$ = 1/2 and $J_{eff}$ = 3/2 bands may ultimately overlap such that the $J_{eff}$ = 1/2 state is partially filled with electrons and the $J_{eff}$ = 3/2 contains a corresponding number of holes. This may result in a magnetic moment.

(2) The exchange $J_H$ and direct Coulomb U interactions couple different orbitals. Recent theoretical studies [15, 29-32] suggest increased electron hopping promotes a quantum phase transition from the expected $J_{eff}$ = 0 state to a novel magnetic state with local moments on the $5d^4$ ions.

(3) A band structure study of a series of double-perovskite iridates with $Ir^{5+}(5d^4)$ ions confirms the existence of a magnetic state in distorted $Sr_2YIrO_6$ [28], and predicts a breakdown of the $J_{eff}$ = 0 state in undistorted, cubic $Ba_2YIrO_6$, as well. This approach attributes magnetic order to a band structure effect, rather than non-cubic crystal effects, in these double-perovskites [31]. However, a more recent theoretical study predicts a non-magnetic state in double-perovskite iridates [33]. No long-range magnetic order was detected in recent experimental studies on $Ba_2YIrO_6$ [34-36], although correlated magnetic moments (0.44 $\mu_B$/Ir) are detected below 0.4 K in one of these studies [36]. The other two studies did not investigate properties below 2 K, which is essential for determining the nature of the magnetic ground state [34, 35].

The stability limits of spin-orbit-coupled $J_{eff}$ states in heavy transition metal materials urgently require more careful investigation. Double-perovskite iridates with a face centered cubic (FCC) structure **[37, 38]** is a class of materials particularly interesting in this regard; the elongated separation of the transition metal ions suppresses direct electron hopping, and further enhances the effect of the SOI, which in turn promotes quantum fluctuations and, thus, exotic phases in double perovskites with $d^1$ and $d^2$ ions **[39, 40]**.

Herein, we report a magnetic ground state in single-crystal double-perovskite Ba$_2$YIrO$_6$ and Sr-doped Ba$_2$YIrO$_6$ having Ir$^{5+}$($5d^4$) ions. A magnetic transition is observed in both DC and AC susceptibility data near T$_N$, which is below 1.7 K. The AC susceptibility is a particularly sensitive tool for probing nonlinear behavior near magnetic phase transitions in relatively weak magnetic materials such as iridates. Indeed, the heat capacity exhibits a peak near T$_N$, further confirming the existence of the long-range, bulk magnetic order. The ordered state is highly sensitive to even low magnetic fields, which re-affirms a vulnerability of the ground state to small changes in system parameters. The observed properties of the magnetic state cannot be explained by either the low-spin S = 1 state in materials with $d^4$ ions, or the J$_{eff}$ = 0 singlet state, and it instead appears that the ground state is situated somewhere between them. Data presented herein, along with those published earlier **[28]** for Sr$_2$YIrO$_6$ yield a phase diagram that features a linear dependence of T$_N$ on Sr doping, and imply a common mechanism driving the magnetic ground state across the entire series of (Ba$_{1-x}$Sr$_x$)$_2$YIrO$_6$ compounds. Moreover, our study suggests structural distortions are relevant, but not critical to the existence of the magnetic ground state as was anticipated earlier **[28]**.

The magnetic ground state of Sr$_2$YIrO$_6$ we observe is extraordinary because it occurs in the face of two strongly unfavorable conditions: 1) A strong SOI, and 2) quantum fluctuations due to geometric frustration, both favor a singlet $J_{eff} = 0$ ground state. The emergence of magnetic order throughout the entire series of $(Ba_{1-x}Sr_x)_2YIrO_6$ compositions under such highly unfavorable circumstances provides strong evidence for the breakdown of the expected singlet $J_{eff} = 0$ state. The SOI is expected to be relatively enhanced in the double-perovskite structure. We infer a highly delicate nature of the magnetic state, which argues for a subtle balance existing between the SOI, U, Hund's rule correlations and electron hopping.

## II. Experimental details

Single crystals of $Ba_2YIrO_6$ and Sr doped $Ba_2YIrO_6$ were grown using a self-flux method from off-stoichiometric quantities of $IrO_2$, $BaCO_3$ (and $SrCO_3$) and $Y_2O_3$. The mixed powders were fired to 1440 °C for 5 hours and then slowly cooled at a rate of 2 °C/hour. The obtained single crystals had typical dimensions 1.0 x 1.0 x 0.5 mm$^3$, and the crystal structures of $Ba_2YIrO_6$ and $(Ba_{0.63}Sr_{0.37})_2YIrO_6$ were determined using a Rigaku XtaLAB PRO X-ray diffractometer equipped with a PILATUS 200K hybrid pixel array detector at the Oak Ridge National Laboratory. Full data sets for more than 20 crystals were collected at 100 K and the structures were refined using SHELX-97 and FullProf software [41, 42] (see **Tables 1** and **2**). Chemical compositions were determined using both energy dispersive X-ray analysis (EDX) (Hitachi/Oxford 3000) and single-crystal X-ray diffraction.

Measurements of DC magnetic susceptibility and magnetization were carried out using a Quantum Design MPMS-7 SQUID Magnetometer equipped with helium-3 refrigeration capable of reaching low temperatures down to 0.44 K. A Quantum Design

Dynacool PPMS System equipped with a dilution refrigerator and a 14-Tesla magnet enabled AC magnetic susceptibility measurements over the temperature range 0.05 K- 4 K using small AC drive fields (< 4 Oe rms). The PPMS was also used to perform heat capacity measurements in applied magnetic fields up to 14 T and temperatures down to 0.05 K. We stress that access to low temperatures was essential to obtain adequate characterization of the magnetic ground state of the double-perovskite iridate samples.

**Table 1**. Structural parameters for single-crystal $Ba_2YIrO_6$ with *Fm-3m* (No. 225)

| T = 100 K | **$a$ = 8.348 (5) Å, V = 581.85 (78) Å³**. The agreement factor $R_1$ = 3.72% was achieved by using 95 unique reflections with I > 4σ and resolution of $d_{min}$ = 0.65 Å. Anisotropic atomic displacement parameters were used for all elements. | | | | | |
|---|---|---|---|---|---|---|
| | Site | x | y | z | Occupancy | $U_{eq}(Å^2)$ |
| Ba | 8c | 0.25 | 0.25 | 0.25 | 1 | 0.0063(5) |
| Y | 4a | 0 | 0 | 0 | 1 | 0.0049(6) |
| Ir | 4b | 0.5 | 0.5 | 0.5 | 1 | 0.0059(4) |
| O | 24e | 0.2638(11) | 0 | 0 | 1 | 0.011(2) |

**Table 2**. Structural parameters for $(Ba_{0.63}Sr_{0.37})_2YIrO_6$ with *Fm-3m* (No. 225)

| T = 100 K | **$a$ = 8.280 (4) Å, V = 567.74 (75) Å³**. The agreement factor $R_1$ = 5.43% was achieved by using 90 unique reflections with I > 4σ. | | | | | |
|---|---|---|---|---|---|---|
| | Site | x | y | z | Occupancy | $U_{eq}(Å^2)$ |
| Ba | 8c | 0.25 | 0.25 | 0.25 | 0.63(2) | 0.0072(4) |
| Sr | 8c | 0.25 | 0.25 | 0.25 | 0.37(2) | 0.0072(4) |
| Y | 4a | 0 | 0 | 0 | 1 | 0.0049(4) |
| Ir | 4b | 0.5 | 0.5 | 0.5 | 1 | 0.0056(3) |
| O | 24e | 0.2632(7) | 0 | 0 | 1 | 0.027(2) |

We also emphasize that the AC magnetic susceptibility χ(T) can be extremely sensitive to thermodynamic phase changes since it is a differential susceptibility defined as *d*M/*d*H, and measures slope changes in the magnetization M(H) (H = magnetic field) rather than the magnitude of the magnetization itself. The AC susceptibility is therefore a particularly sensitive tool for probing nonlinear behavior near magnetic phase transitions in relatively weak magnetic materials such as iridates. The low-temperature AC susceptibility data (for 0.05 K– 4 K) offer additional confirmation of the existence of magnetic order in the double-perovskite iridates, and provide valuable clarification of the previous conflicting experimental and theoretical reports.

### III. Results and discussion

Both $Ba_2YIrO_6$ and $(Ba_{0.63}Sr_{0.37})_2YIrO_6$ adopt a cubic structure with space group *Fm-3m* (No. 225) in which every other Ir is replaced by nonmagnetic Y and the remaining magnetic $Ir^{5+}$ ions form a network of edge-sharing tetrahedra arranged in a FCC structure **[34, 37, 38]**, as shown in **Fig. 1**, and **Tables 1** and **2**. No appreciable intersite disorder is observed because of the significant differences in oxidation state and ionic radii between $Y^{3+}$ and $Ir^{5+}$ ions (the ionic radius $r(Y^{3+})$ = 0.90 Å, and $r(Ir^{5+})$ = 0.57 Å). Indeed, the structure refinement cannot converge at all when we introduce Y at Ir sites or Ir at Y sites. It is emphasized that this result rules out any possible intermixing of Y and Ir. The lattice parameters of $Ba_2YIrO_6$ and $(Ba_{0.63}Sr_{0.37})_2YIrO_6$ as a function of temperature from 100 K to 300 K are presented in **Figs. 1a** and **1b** and **Tables 1** and **2**. The lattice parameters decrease with Sr doping as expected, but change only slightly with temperature, especially in the case of $Ba_2YIrO_6$. More importantly, the crystal structure of both compounds is cubic, with no discernable distortions, which rules out nuclear Schottky effects in zero magnetic field at low temperatures. The stable, symmetric structure of the Ba double perovskites contrasts with the crystal structure of $Sr_2YIrO_6$ in which the $IrO_6$ octahedra are tilted and rotated, causing flattened $IrO_6$ octahedra, which generates a non-cubic crystal field that has important consequences for physical properties **[28]**.

Ba$_2$YIrO$_6$ displays paramagnetic behavior at temperatures above 1.7 K. The DC magnetic susceptibility χ(T) at μ$_o$H=1 T, defined as M/H follows a Curie-Weiss law for 50 < T < 300 K, as shown in **Fig. 2a**. (Note that the isothermal magnetization M(H) is linear at T > 1.8 K and μ$_o$H < 5 T, therefore, χ(T) = ΔM/ΔH =M/H.) Data fits to the Curie-Weiss law over the range 50 < T < 300 K yield an effective moment μ$_{eff}$ = 1.44 μ$_B$/Ir and a Curie-Weiss temperature θ$_{CW}$ = -149 K. The value of μ$_{eff}$ is considerably smaller than the value 2.83 μ$_B$/Ir expected for a spin-only S = 1 state with *5d$^4$* ions. A reduced value of μ$_{eff}$ is commonplace in iridates, in part because the strong SOI causes a partial cancellation of the spin and orbital contributions **[15, 26]**. A strong AFM exchange coupling may be deduced from the large magnitude of θ$_{CW}$; however, the absence of any magnetic order at T > 1.7 K suggests the existence of strong quantum fluctuations, which can be interpreted as evidence for a strong SOI in the geometrically frustrated FCC lattice of Ba$_2$YIrO$_6$. A signature for long-range magnetic order is evident at T$_N$ = 1.48 K, as shown in **Fig. 2b**. The two temperature scales evident in the magnetic data yield a strikingly large frustration parameter, |θ$_{CW}$|/T$_N$ = 97.3, which apparently reflects the strong depression of long-range magnetic order observed in this system. The isothermal magnetization M(H) for T = 0.5 K initially rises and then exhibits a weak slope change near 4.5 T. This slope change disappears at T = 0.8 K (see **Fig. 2c**). Note that M(H) at T = 0.5 K is weaker than M(H) at T = 0.8 K below μ$_o$H = 5 T, suggesting AFM order strengthens at lower temperatures. Furthermore, there is evidence of another magnetic anomaly near 0.6 K (**Fig. 2b**), which is also supported in the specific heat C(T) data, as discussed below. It is noteworthy that the values of μ$_{eff}$ and θ$_{CW}$ which we obtained are different from those reported in another experimental study **[36]**; these differences are likely due to different temperature ranges adopted in the Curie-Weis law fitting. In general, Curie-Weiss fits that include higher temperatures are more desirable, because relatively large exchange interactions are extant in the iridates.

Similar magnetic behavior is observed in $(Ba_{0.63}Sr_{0.37})_2YIrO_6$, as shown in **Fig. 3**. However, best-fit values $\theta_{CW}$ = -18 K, and $\mu_{eff}$ = 0.64 $\mu_B$/Ir, are significantly lower than those deduced for $Ba_2YIrO_6$ (**Fig. 3a**), which may be attributed to atomic disorder —"order out of disorder" which reduces the effects of frustration. The onset of $T_N$ is also lower than that for the pure compound (**Fig. 3b**). M(H) exhibits a stronger slope change (**Fig. 3c)** that mimics that seen in $Sr_2YIrO_6$ (where M(H) initially rises, then exhibits a plateau, and is followed by a rapid jump at a critical field [**28**]).

The evidence for long-range magnetic order seen in **Figs. 2** and **3** is confirmed by measurements of the specific heat C(T) in which an onset of a peak occurs near $T_N$ for both $Ba_2YIrO_6$ and $(Ba_{0.63}Sr_{0.37})_2YIrO_6$, as shown in **Figs. 4a** and **4c**, respectively. The peak of C(T) completes below 1 K, where a possible second magnetic anomaly is observed in $\chi(T)$, particularly for $Ba_2YIrO_6$ (**Fig. 2b**). A nuclear Schottky contribution may be unlikely because in zero magnetic field it requires a non-cubic crystal electric field to generate the necessary electric field gradient and quadrapole splitting; both of the compounds adopt a cubic structure without discernable distortions. The entropy removal due to the magnetic transition is only a tiny fraction of Rln(2S+1) = Rln(3) = 9.13 J/mole K expected for a magnetic transition of an S = 1 system. The measured S(T) is approximately 0.07 J/mole K for $Ba_2YIrO_6$ at 4 K (**Fig. 4b;** note a more rapid drop in S(T) below 0.4 K) and 0.08 J/mole K for $(Ba_{0.63}Sr_{0.37})_2YIrO_6$ at 4 K (**Fig. 4d**). S(T) for both compounds is similar to but greater than the 0.025 J/mole K for $Sr_2YIrO_6$ [**28**]. The small entropy removal by the low-temperature peaks in C(T) seems to be a signature of this group of double-perovskite iridates, consistent with the fragile nature of the magnetic order [28]. Indeed, the magnetic ground state is delicate so that even moderate fields are strong enough to produce significant changes in C(T). This qualitatively explains the strongly depressed magnetic order (**Figs. 2** and **3**) and drastically enhanced magnetic anomaly in C(T) at $\mu_oH = 2$ T for $Ba_2YIrO_6$ and at $\mu_oH = 1.5$ T for $(Ba_{0.63}Sr_{0.37})_2YIrO_6$ (**Figs. 4a** and **4c**). Remarkably, the magnetic order is not at all suppressed in magnetic fields, which sharply contrasts with conventional AFM order. It deserves to be mentioned that a small induced moment in an otherwise nonmagnetic ground state could be due to a van Vleck interaction, which is the second-order Zeeman effect. Therefore, the moment could be very sensitive to thermal and magnetic parameters.

The magnetic ordering transitions of $Ba_2YIrO_6$, $(Ba_{0.63}Sr_{0.37})_2YIrO_6$ and $Sr_2YIrO_6$ have been thoroughly examined over the temperature range 0.05 K - 4 K. Given the high sensitivity of AC magnetometry to magnetic phase transitions, we have acquired data for AC $\chi(T)$ as a useful probe for the existence of an ordered magnetic state, as shown in **Fig. 5**. A pronounced peak in AC $\chi(T)$ defines $T_N$, and confirms the presence of low-temperature magnetic order in each of these pentavalent iridates.

Data for M/H for $Ba_2YIrO_6$, $(Ba_{0.63}Sr_{0.37})_2YIrO_6$ and $Sr_2YIrO_6$ are available in **Fig. 6a** for comparison to the AC $\chi(T)$ data shown in **Fig. 5**, and illustrate the evolution of the magnetic behavior in the series of compositions $(Ba_{1-x}Sr_x)_2YIrO_6$. It is important to note that all results of M/H, AC $\chi$ and C(T) show that $T_N$ decreases linearly with increasing Sr concentration although $T_N$ occurs at a slightly different temperature for each compound (**Fig. 6c**). Such small differences in $T_N$ are expected, given the different nature of the experimental probes. Moreover, the pronounced metamagnetic transition that occurs in $Sr_2YIrO_6$ is significantly weakened in $Ba_2YIrO_6$ (**Fig. 6b**). The metamagnetic transition, which is a signature of a spin reorientation, appears to be closely associated with structural distortions that are strong in $Sr_2YIrO_6$. It is thus not surprising that the magnetization M systematically decreases with x. A representative value of M at 7 T and 0.5 K as a function of x is presented in **Fig. 6d**. Note that both $T_N$ and M(7T) share a similar x-dependence (**Figs.6c and 6d**).

Our study indicates the existence of a magnetically ordered ground state in the double-perovskite iridates. However, the small value of the ordered magnetic moment and small magnetic entropy removal associated with heat capacity anomalies imply that this magnetic state is weak and barely stable compared to either the S = 1 state that is commonly observed among heavy $d^4$ transition element ions, or the $J_{eff}$ = 0 state driven by strong SOI. These circumstances make this magnetic state unique and intriguing. Indeed, several theoretical studies predict a quantum phase transition from the nonmagnetic $J_{eff}$ = 0 state to a magnetic state **[15, 29-32]**. A more commonly accepted argument is that the SOI competes with a comparable exchange interaction $J_H$ and a generically large electron hopping (because of the extended nature of 5d-orbitals) that suppresses the singlet $J_{eff}$ = 0 ground state. At the same time, the SOI breaks down the low-spin S = 1 state. This "simultaneous" destabilization of the S = 1 and $J_{eff}$ = 0 states leads to a magnetic state that resides somewhere between the other two.

The above picture is not without controversy. A band structure study claims a non-magnetic state having a spin gap of 200 meV occurs in $Ir^{5+}$ double perovskites **[33]**. However, a gap of this magnitude implies the magnetic susceptibility should not have any significant temperature dependence at low temperatures, which is in apparent contradiction with experimental data **[33-36]**.

## IV. Summary

While the $J_{eff} = 1/2$ insulating state model successfully captures new physics observed in many iridates, recent studies suggest that it may not be adequate to describe new phenomena that are observed when the relative strength of the SOI critically competes *with both* the strength of electron hopping and exchange interactions. It is worth mentioning that the $J_{eff} = 1/2$ model takes a single-particle approach that assumes that Hund's rule coupling among the electrons can be neglected. This assumption needs to be closely examined when the exchange correlations underlying Hund's rules are no longer negligible. Nevertheless, the observation of a magnetic ground state in the double perovskite iridates may indeed imply that the SOI is not as dominant as generally assumed. Moreover, the stability of magnetic order in these situations could be extraordinarily fragile, as evidenced in this study, and indicated by the varied magnetic behavior reported in other recent experimental and theoretical studies **[28-36]**. It is clear that the stability limits of the spin-orbit-coupled $J_{eff}$ states in heavy transition metal materials must be further investigated.

## Acknowledgments

GC is indebted to Drs. Ribhu Kaul, Arun Paramekanti, Tanusri Saha-Dasgupta, Nandini Trivedi, Roser Valenti, L.T. Corredor and B. Buchner for stimulating discussions. This work was supported by NSF grants DMR-1265162 and DMR-1712101 (GC), DMR-1506979 (LED), and Department of Energy (BES) grant No. DE-FG02-98ER45707 (PS).

*Email: gang.cao@colorado.edu*

**Captions:**

**Fig. 1.** *Upper panel*: The double-perovskite crystal structure of $Ba_2YIrO_6$ based on the single-crystal diffraction data (Green: Ba, Yellow: Ir, Blue: Y, and red: Oxygen). *Lower panel*: Temperature dependence of **(a)** the lattice parameters and **(b)** the oxygen coordinate $(x,0,0)$ for $Ba_2YIrO_6$, and $(Ba_{0.63}Sr_{0.37})_2YIrO_6$. Both systems remain in the *Fm-3m* space group at all temperature studied.

**Fig. 2.** Magnetic properties of $Ba_2YIrO_6$: The temperature dependence of **(a)** the DC magnetic susceptibility $\chi(T)$ (left scale) and $1/\Delta\chi$ (right scale) at $\mu_oH = 1$ T for $1.7$ K $\leq$ T $\leq 300$ K (Note that $\Delta\chi=\chi-\chi_o$, where $\chi_o$ is a temperature-independent contribution to $\chi$), and **(b)** the low-temperature magnetization M at $\mu_oH = 1$ T for $0.44$ K $\leq$ T $\leq 4$ K; **(c)** The isothermal magnetization M(H) at T=0.5 and 0.8 K.

**Fig. 3.** Magnetic properties of $(Ba_{0.67}Sr_{0.37})_2YIrO_6$: The temperature dependence of **(a)** the DC $\chi(T)$ (left scale) and $1/\Delta\chi$ (right scale) at $\mu_oH = 1$ T for $1.7$ K $\leq$ T $\leq 300$ K, and **(b)** the low-temperature M at $\mu_oH = 2$ T for $0.44$ K $\leq$ T $\leq 4$ K; **(c)** M(H) at T=0.5 and 0.8 K.

**Fig. 4.** Thermal properties of $Ba_2YIrO_6$ and $(Ba_{0.67}Sr_{0.37})_2YIrO_6$: For $0.05$ K $\leq$ T $\leq 4$ K, the temperature dependence of **(a)** the specific heat C(T) at zero field $\mu_oH = 0$ T and 2 T (note that C (0.1K, 2T) is still finite, 0.002 J/mole K), and **(b)** the entropy removal S(T) at $\mu_oH = 0$ for $Ba_2YIrO_6$; **(c)** C(T) at $\mu_oH = 0$ and 1.5 T and **(d)** S(T) at $\mu_oH = 0$ for $(Ba_{0.67}Sr_{0.37})_2YIrO_6$. Insets in **(b)** and **(d)**: dC/dT vs T for $Ba_2YIrO_6$ and $(Ba_{0.67}Sr_{0.37})_2YIrO_6$, respectively.

**Fig. 5.** The temperature dependence of the AC $\chi(T)$ for **(a)** $Ba_2YIrO_6$, **(b)** $(Ba_{0.67}Sr_{0.37})_2YIrO_6$ and **(c)** $Sr_2YIrO_6$. The AC field is 3 Oe and AC frequency is 5 kHz.

**Fig. 6.** Comparison of magnetic behavior for the series of $(Ba_{1-x}Sr_x)_2YIrO_6$: **(a)** The temperature dependence of M/H ($\mu_oH=1$ T for x=0 and 1 and 2T for x=0.37); **(b)** The isothermal M(H) at T=0.5 K; **(c)** The magnetic transition $T_N$ as a function of Sr doping x obtained from the data of M(T), AC $\chi$(T) and C(T); **(d)** The value of the magnetization M at 7 T and 0.5 K as a function of Sr doping x.

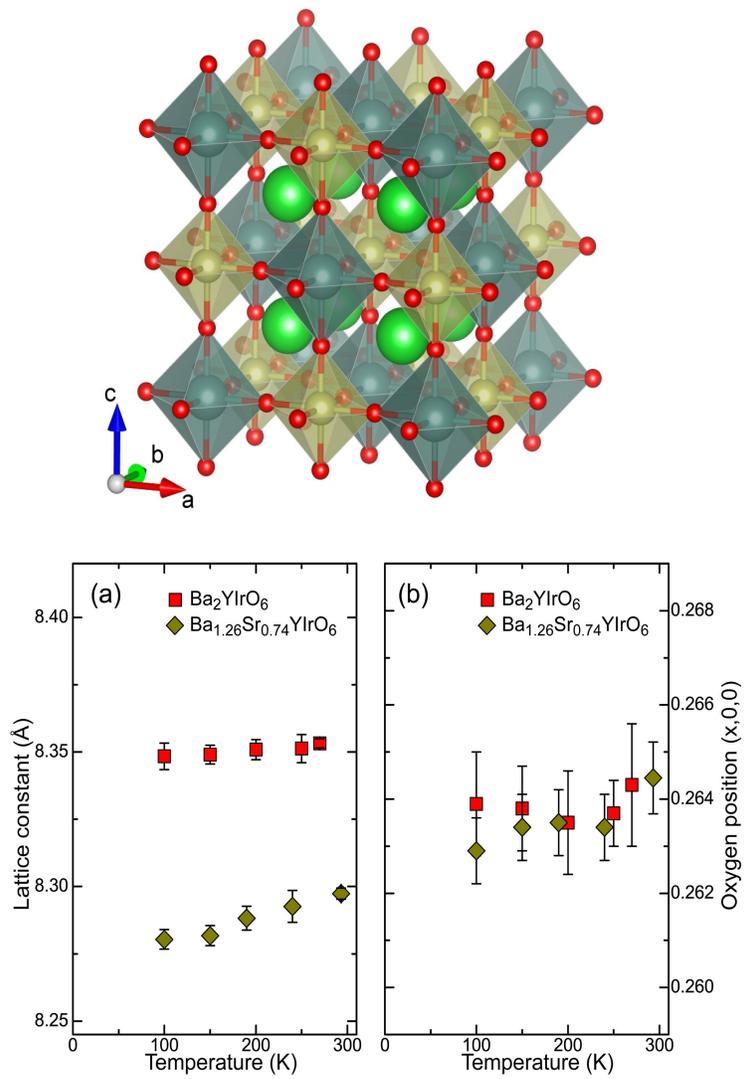

Fig. 1

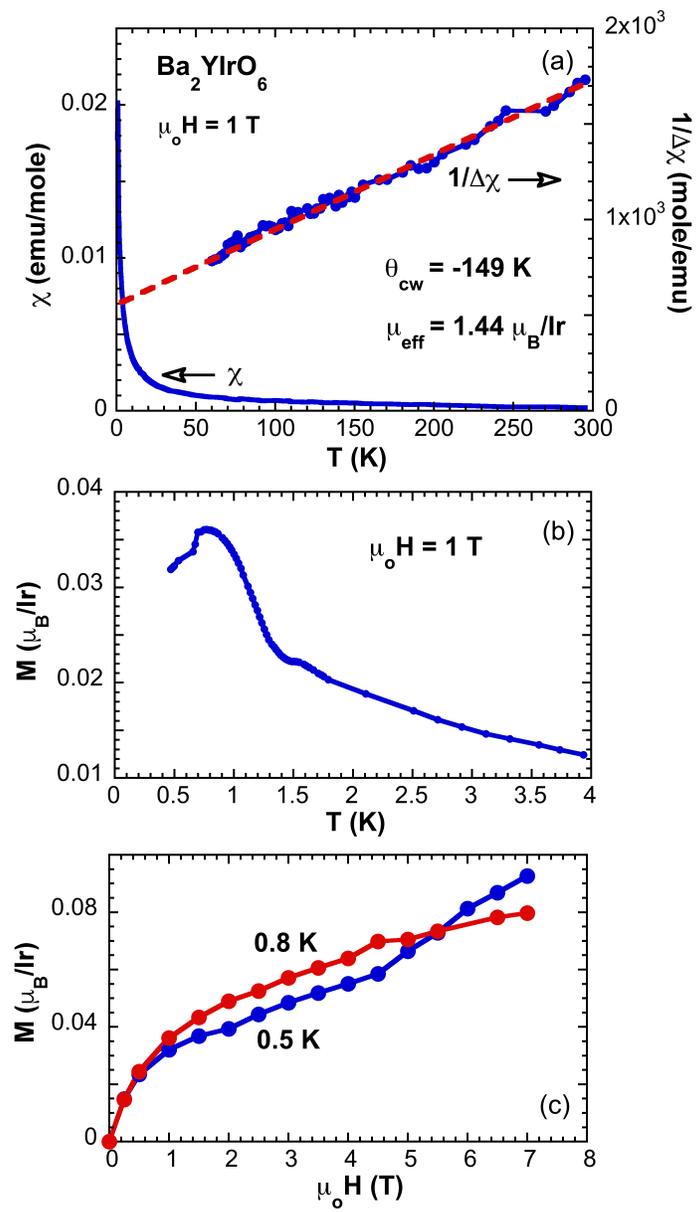

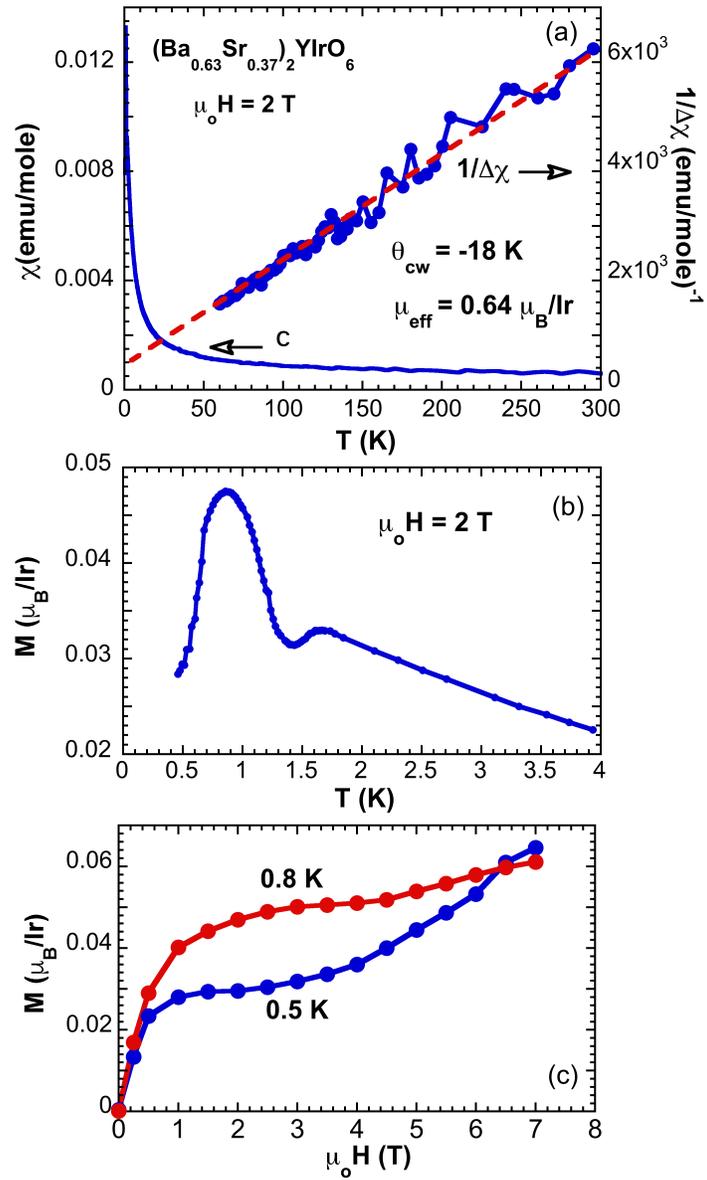

Fig.3.

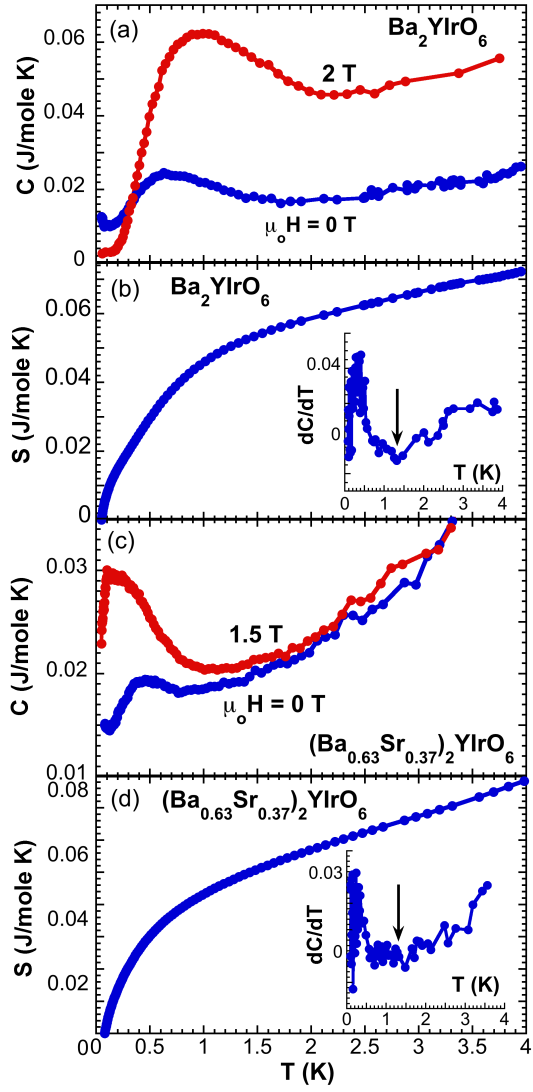

Fig.4

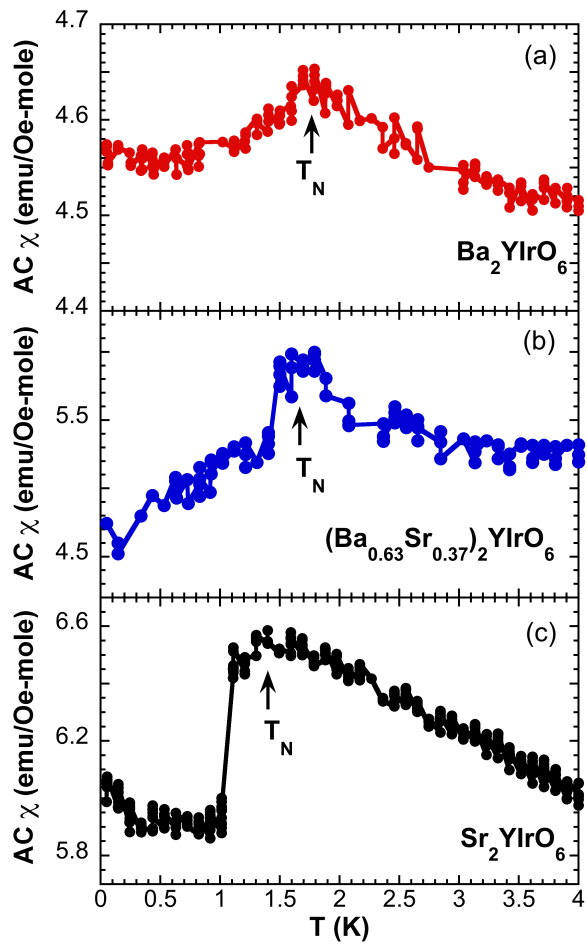

Fig.5

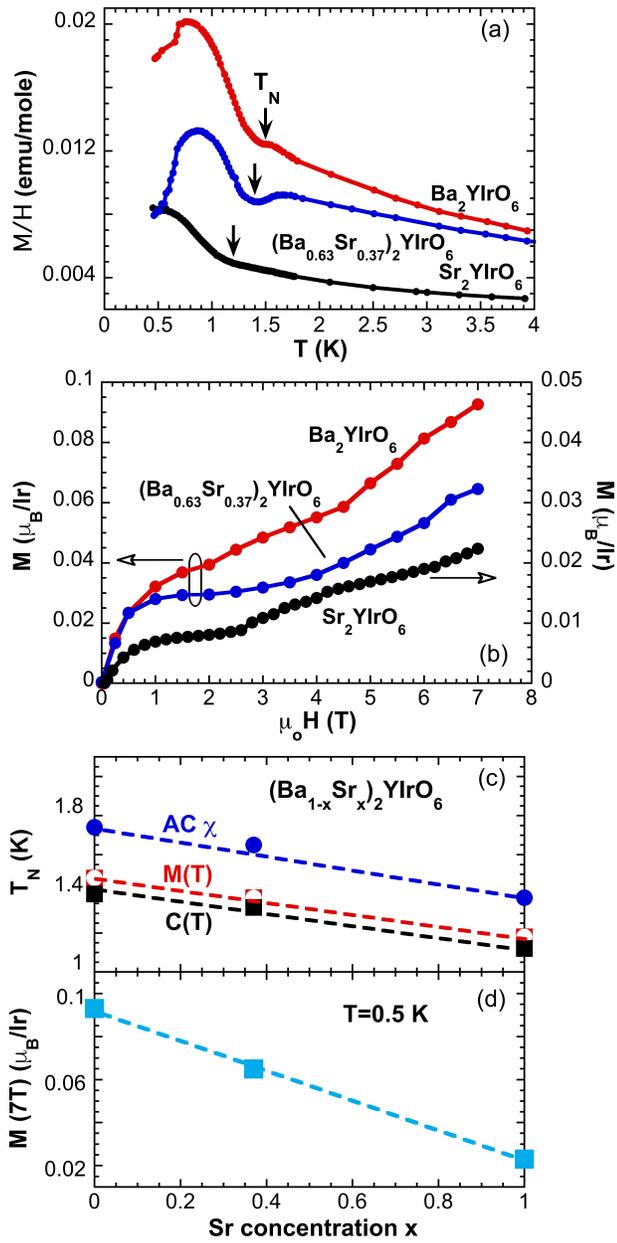

Fig. 6